# Comparative Study of Acoustic Echo Cancellation Algorithms for Speech Recognition System in Noisy Environment

Urmila Shrawankar

RTM Nagpur University, Nagpur (MS), INDIA

Email : urmila@ieee.org

*Abstract :*
Traditionally, adaptive filters have been deployed to achieve AEC by estimating the acoustic echo response using algorithms such as the Normalized Least-Mean-Square (NLMS) algorithm. Several approaches have been proposed over recent years to improve the performance of the standard NLMS algorithm in various ways for AEC. These include algorithms based on Time Domain, Frequency Domain, Fourier Transform, Wavelet Transform Adaptive Schemes, Proportionate Schemes,
Proportionate Adaptive Filters, Combination Schemes, Block Based Combination, Sub band Adaptive Filtering, Uniform Over Sampled DFT Filter Banks, Sub band Over-Sampled DFT Filter
Banks, Volterra Filters, Variable Step-Size (VSS) algorithms, Data Reusing Techniques, Partial Update Adaptive Filtering Techniques and Sub band (SAF) Schemes. These approaches aim to address issues in echo cancellation including the performance with noisy input signals, Time-Varying echo paths and computational complexity. In contrast to these approaches, Sparse Adaptive algorithms have been developed specifically to address the performance of adaptive filters in sparse system identification.
In this paper we have discussed some AEC algorithms followed by comparative study with respective to step-size, convergence and performance.

## I. INTRODUCTION

The echo is produced when the far-end signal activates a (possibly time-varying) echo path. This echo signal is superimposed upon the near-end signal, which is possibly contaminated by additive noise [1, 2]. The goal of an echo canceller is to produce a replica of the echo signal which can be used to remove the echo before the signal is delivered to the far-end.

Based on the approach acoustic echo, seriously degrades user experience due to the coupling between the loudspeaker and microphone. The time variation of the near-end acoustic impulse response (AIR) may arise due to, a change in temperature [20], pressure and changes in the acoustic environment. It is also well known that the reverberation time of an AIR is proportional to the volume of the enclosed space and inversely proportional to the absorption area [21]. For an outdoor environment, the reverberation time is reduced significantly due to the lack of reflections from any enclosure.

The main part of the echo cancellation application can be interpreted as a system identification problem, where an adaptive filter [13,14] is used to identify an unknown system, i.e., the echo path [37]. However, the echo paths for acoustic echo cancellation scenarios are sparse in nature, i.e., a small percentage of the impulse response components have a significant magnitude while the rest are very small or even zero.

The degree of sparsity [1] of a particular echo path is typically not known beforehand, and it can even be time varying. For this reason, it is desirable to develop schemes that shows robust performance to different levels of sparseness in the echo channel.

In the context of acoustic echo cancellation (AEC), it is shown that the level of sparseness in acoustic impulse responses can vary greatly in a mobile environment. When the response is strongly sparse, convergence of conventional approaches is poor.

The common idea of this kind of algorithms is to update each coefficient of the filter independently of the others, by adjusting the adaptation step-size in proportion to the magnitude of the estimated filter coefficient to derive proportionate-type APAs in a more rigorous manner [11, 12].

This paper explains varies acoustic echo cancellation schemes and algorithms in section II, Proportionate-Type Algorithms for Echo Cancellation followed by comparative study with respect to step-size, convergence and performance in section III and in section IV, Sparseness Controlled Algorithms are discussed with their comparison finally conclusion is given in section V.

## II. CLASSIFICATION OF ACOUSTIC ECHO CANCELLATION (AEC) SCHEMES AND ALGORITHMS

### A. Adaptive Schemes

It is a well-known fact that adaptive schemes which distribute the adaptation energy equally among all filter coefficients, such as Least-Mean-Square (LMS) and Normalized LMS (NLMS), exhibit a very slow convergence for filters [22,23], Makino et al. [24] introduced the exponentially-weighted step size NLMS(ES-NLMS), which assigns a different adaptation speed to each coefficient of the echo canceller the PNLMS algorithm tries to accelerate the convergence of the filter by adapting faster the weights corresponding to the active region of the sparse echo path. The advantage of PNLMS over ES-NLMS is that it does not assume any other a priori knowledge about the echo channel.

### B. Proportionate Schemes

Proportionate Schemes [2] offer better behavior than standard adaptive filters for the cancellation of sparse echopath impulse responses, but the selection of their parameters is subject to different compromises.

Proportionate Schemes offer improved performance when the echo path is sparse, but are still subject to some compromises regarding their convergence properties and steady-state error.

### C. Proportionate Adaptive Filters

Proportionate Adaptive Filters, such as those based on the Improved Proportionate Normalized Least-Mean-Square (IPNLMS) algorithm, have been proposed for echo cancellation as an interesting alternative to the Normalized Least-Mean-Square (NLMS) filter.

The adaptive combination of adaptive filters has proved to be an effective approach to improve the performance of adaptive schemes and to simplify their use. The basic idea of the algorithms [26,27] is to combine adaptive filters with different settings, so that the combination behaves, at each iteration, as the best component filter [28].

### D. Block-Based Combination

A new block-based combination has also been introduced to further reduce the steady-state misadjustment. The performance of such combination schemes has been illustrated when combining IPNLMS filters with different settings, showing in all cases significant advantages over the use of a single filter.

E. **Combination Schemes**

Combination schemes can help to improve the performance of proportionate filters, by alleviating the speed versus precision tradeoff, as well as by increasing robustness to channels with different degrees of sparsity. the combination can achieve better-than-universal performance.

F. **Subband Adaptive Filtering**

Among different techniques for AEC, subband adaptive filtering [15] has the advantage of high computational efficiency and fast rate of convergence [29].

Due to the nature of subband adaptive filtering, critically sampled FBs are generally considered unsuitable, despite of their attractiveness and successful applications in compression. In [30] it was shown that if critically sampled FBs are to be used correctly for adaptive filtering, one must consider cross-band filters, at the cost of significantly increased computation complexity.

In this paper, we focus on the design of NPR oversampled uniform filter banks using discrete Fourier transform (DFT) [7,8] modulation.

G. **Uniform Over-Sampled DFT Filter Banks**

When viewed from the bandpass filter interpretation, the FBs first process the input signal $x(n)$, by K bandpass filters denoted by $h_k(n)$. The resulting signals are decimated by a factor of M and become $X_k(m)$. With AEC, each of the K channels has an adaptive filter that minimizes the echo energy, resulting in echo cancellation [29].

H. **Subband Over-Sampled DFT filter banks (FB) [3]**

The design requirements typically are good echo cancellation quality, low delay, small reconstruction error, and low computation complexity. This method explicitly includes a model for Echo Return Loss Enhancement (ERLE) as part of the optimization criteria. Convergence of the high-dimensional, nonlinear optimization problem is facilitated by decorrelating the prototype filter impulse response via a Discreet Cosine Transform (DCT), discarding many insignificant coefficients and thus reducing the dimensionality of the search. The current research shows the effectiveness of design and the effects on the performance of AEC, with ELRE improvements on the order of 3 dB or better.

I. **Time-Domain Considerations**

While the frequency domain characteristics are necessary for designing a good prototype filter, they are not sufficient. For example, one may satisfy the frequency domain criteria by relying mostly on the time domain aliasing cancellation. However, because the gain factors for each subband are subject to independent changes over time as a result of AEC, time domain cancellation becomes much less effective. Therefore, it is reasonable that a good solution should strike a balance between time and frequency domain criteria.

J. **Volterra filters : Improved Nonlinear System Identification in the Short-Time Fourier Transform (STFT) Domain.**

An important property of Volterra filters [9,10], which makes them useful in nonlinear estimation problems, is the linear relation between the system output and the filter coefficients. Many approaches, which attempt to estimate the Volterra kernels in the time domain, employ conventional linear estimation methods in batch [32, 33] or adaptive forms [4, 31, 34].

A common difficulty associated with time-domain methods is their high computational cost, which is attributable to the large number of parameters of the Volterra model. This problem becomes even more crucial when estimating systems with relatively large memory length, as in acoustic echo cancellation applications.

A Discrete Frequency-Domain model [35,36], is defined, which approximates the Volterra filter in the frequency domain using multiplicative terms.

## III. PROPORTIONATE-TYPE ALGORITHMS FOR ECHO CANCELLATION

Proportionate-type Normalized Least-Mean-Square algorithms are developed in the context of echo cancellation. In order to further increase the convergence rate and tracking, the "proportionate" idea is applied to the Affine Projection Algorithm (APA) [5] in a straightforward manner. These algorithms converge faster than the Normalized Least-MeanSquare (NLMS) Algorithm, especially for sparse impulse responses. It is clear that the NLMS algorithm is obtained when all the coefficients receive the same increment. These parameters distinguish between different proportionate-type NLMS algorithms. Among these, the IPNLMS algorithm [38] is one of the most attractive choices, mainly because of its robustness to echo paths with different sparseness degrees.

A. **NLMS :** The NLMS algorithm is one of the most popular for AEC due to its straightforward implementation and low complexity compared to, other algorithms, for example, the Recursive Least Squares (RLS) algorithm.
B. **PNLMS :** The PNLMS (proportionate NLMS) [6] and MPNLMS (µ PNLMS ) [19] algorithms have been proposed for sparse system identification. It prevents the coefficients from stalling when they are much smaller than the largest coefficient.
C. **MPNLMS : The MPNLMS ( µ PNLMS)** algorithm is proposed to improve the convergence of PNLMS
D. **IPNLMS :** The IPNLMS (Improved PNLMS) [17,18] algorithm was originally developed for NEC and was further developed for the identification of acoustic room impulse responses. It employs a combination of proportionate (PNLMS) and non-proportionate (NLMS) adaptation. IPNLMS [25] is a very good approximation of the filter, while being more convenient from a practical point of view.

A. **Comparative Study of Echo Cancellation Algorithms**

| *Comparative Study With Respective To Step-Size* | |
|---|---|
| **NLMS** | • Step-size is the same for all filter coefficients, |
| **PNLMS** | • PNLMS employs larger step-sizes for active coefficients than for inactive coefficients and consequently achieves faster convergence than NLMS for sparse impulse responses.<br>• PNLMS suffers from slow convergence giving rise to a small step-size for each large coefficient.<br>• The PNLMS algorithm achieves a high rate of convergence by employing stepsizes that are proportional to the magnitude of the estimated impulse response coefficients |
| **MPNLMS** | • MPNLMS suffer from slow convergence giving rise to a small step-size for each large coefficient. This causes a significant degradation in convergence performance for PNLMS and MPNLMS when the impulse response is dispersive such as can occur in AIRs.<br>• It achieves convergence by computing the optimal proportionate step-size during the adaptation process. |

| IPNLMS | • A large step size results in faster convergence, while the residual misadjustment is reduced for small. |
|---|---|
| *Comparative Study With Respective To Convergence* | |
| NLMS | Convergence rate reduces significantly when the impulse response is sparse |
| PNLMS | • PNLMS achieves fast initial convergence followed by a slower second phase convergence [19].<br>• After a very fast initial response, PNLMS convergence slows down<br>• Parameter imposes a behavior tradeoff for channels with different degrees of sparsity [25]. For strongly sparse channels the fastest convergence is obtained by the PNLMS filter.<br>• PNLMS suffers from slow convergence giving rise to a small step-size for each large coefficient. This causes a significant degradation in convergence performance for PNLMS and MPNLMS<br>• The PNLMS algorithm achieves a high rate of convergence by employing stepsizes that are proportional to the magnitude of the estimated impulse response coefficients<br>• Both PNLMS and MPNLMS suffer from slow convergence when the unknown system is dispersive [16,17]. |
| MPNLMS | • The MPNLMS algorithm was proposed to improve the convergence of PNLMS. It achieves this by computing the optimal proportionate step-size during the adaptation process.<br>• The MPNLMS algorithm was derived such that all coefficients attain a converged value to within a vicinity of their optimal value in the same number of iterations [19].<br>• Both PNLMS and MPNLMS suffer from slow convergence giving rise to a small step-size for each large coefficient. This causes a significant degradation in convergence performance for PNLMS and MPNLMS |
| IPNLMS | • the IPNLMS algorithm has faster convergence than NLMS and PNLMS regardless of the impulse response nature [18]<br>• In spite of IPNLMS generally achieving faster convergence than NLMS |
| *Comparative Study With Respective To Performance* | |
| NLMS | • NLMS achieves approximately 7-dB better steady-state performance than the MPNLMS |
| PNLMS | • PNLMS behavior degrades significantly when identifying not-so-sparse echo channels |
| MPNLMS | • Gives improved performance then PNLMS |
| IPNLMS | • It does not outperform MPNLMS for highly sparse impulse responses |

## IV. SPARSENESS-CONTROLLED ALGORITHMS FOR ECHO CANCELLATION

A class of AEC algorithms that cannot work well in both sparse and dispersive circumstances, but also adapt dynamically to the level of sparseness using a new sparseness-controlled approach. The sparseness of the AIR of an outdoor environment is significantly greater than typical indoor environments and equally. Variation in the sparseness of AIRs can also occur in AEC within an enclosed space.

In this section, we propose a class of algorithms and their comparison that are robust to the sparseness variation of AIRs. These algorithms compute a sparseness measure of the estimated impulse response at each iteration of the adaptive process and incorporate it into their conventional methods.

- **SC-PNLMS :** The proposed sparseness controlled PNLMS algorithm (SC-PNLMS) inherits the proportionality step-size control over a large range of sparse impulse response. The low values of linear function are allocated for a large range of sparse impulse responses. The algorithm achieves overall desired effect of the robustness to sparse and dispersive AIRs.
- **SC-MPNLMS :** When the estimated AIR is dispersive it follows as in NLMS. the MPNLMS algorithm, the resulting sparseness-controlled MPNLMS algorithm (SCMPNLMS) inherits more of the MPNLMS properties when the estimated AIR is sparse and distributes uniform step-size
- **SC-IPNLMS :** SC-IPNLMS improves the performance of the IPNLMS when the impulse response is sparse, the algorithm allocates more weight to the proportionate term. For sparse AIRs, the algorithm distributes proportionally to the magnitude of the coefficients. As a result of this distribution, the SC-IPNLMS algorithm varies the degree of NLMS and proportionate adaptations according to the nature of the AIRs.

| *Comparative Study with Respective To Step-Size* | |
|---|---|
| The step-size parameter for each algorithm is chosen such that all algorithms achieve the same steady-state [1]. | |
| **SC-PNLMS** | <ul><li>Algorithm Inherits the proportionality step-size control over a large range of sparse impulse response.</li><li>Algorithm also inherits the NLMS adaptation control with larger values this gives a more uniform step-size.</li></ul> |
| **SC-MPNLMS** | <ul><li>Algorithm inherits more of the MPNLMS properties.</li><li>When the estimated AIR is sparse and distributes uniform step-size across.</li><li>When the estimated AIR is dispersive it follows as in NLMS.</li></ul> |
| **SC-IPNLMS** | <ul><li>The step-size control elements for SCIPNLMS estimate different unknown AIRs.</li><li>For dispersive AIRs, algorithm allocates a uniform step-size across.</li><li>For sparse AIRs, the algorithm distributes proportionally to the magnitude of the coefficients. As a result of this distribution, the SC-IPNLMS algorithm varies the degree of NLMS and proportionate adaptations according to the nature of the AIRs.</li></ul> |
| *Comparative Study with Respective To Convergence* [1] | |
| **SC-PNLMS** | <ul><li>When AIR is sparse, SC-PNLMS inherits more of PNLMS properties gives good convergence performance</li><li>When AIR is dispersive, linear function must be small for good convergence performance.</li></ul> |
|   | - The convergence rate of SC-PNLMS is as fast as PNLMS for sparse and much better than PNLMS for dispersive
- SC-PNLMS achieves high rate of convergence similar to PNLMS giving approximately 5-dB improvement in normalized misalignment during initial convergence compared to NLMS for a sparse AIR.
- SC-PNLMS maintains its high convergence rate over NLMS and PNLMS giving approximately 4-dB improvement in normalized misalignment compared to PNLMS for dispersive AIR.
- SC-PNLMS algorithm achieves the highest rate of convergence, giving convergence as fast as PNLMS and approximately 7-dB improvement during initial convergence compared to NLMS for the sparse AIR. For dispersive AIR, SC-PNLMS performs almost the same as NLMS with approximately 4-dB improvement compared to PNLMS.
- SC-PNLMS has a higher rate of convergence for a sparse system compared to a dispersive system. This is due to the initialization choice of linear function, In addition, a smaller value of is favorable for the dispersive AIR, since SC-PNLMS performs similarly to NLMS for small values.
- SC-PNLMS behaves like NLMS, the convergence rate of SC-PNLMS is high when is linear function is small for a dispersive channel.
- The algorithm inherits properties of the NLMS for a small value. the SC-PNLMS algorithm inherits properties of PNLMS giving good performance for sparse AIR before the echo path change. This is because SC-PNLMS inherits the beneficial properties of both PNLMS and NLMS.
- The algorithm achieves overall desired effect of the robustness to sparse and dispersive AIRs. |
| SC-MPNLMS | - SC-MPNLMS algorithm attains approximately 8-dB improvement in normalized misalignment during initial convergence compared to NLMS and same initial performance followed by approximately 2-dB improvement over MPNLMS for the sparse AIR.
- After the echo path change, SCMPNLMS achieves approximately 3-dB improvement compared to MPNLMS and about 8-dB better performance than NLMS for dispersive AIR.
- SC-MPNLMS algorithm achieves approximately 10-dB improvement during initial convergence compared to NLMS and 2 dB compared to MPNLMS for the sparse AIR.
- SC-MPNLMS algorithm achieves an improvement of approximately 4 dB compared to both NLMS and MPNLMS
- By using both WGN and speech input signals, SC-IPNLMS achieves approximately 10-dB improvement in normalized misalignment during initial convergence compared to NLMS for the sparse AIR. |

| SC-IPNLMS | • For comparatively less sparse impulse responses, the algorithm aims to achieve the convergence of NLMS by applying a higher weighting to the NLMS term. |
|---|---|
| | • For a dispersive AIR, the SC-IPNLMS achieves a 5-dB improvement compared to NLMS. For a speech input. |
| | • The improvement of SC-IPNLMS over IPNLMS is 3 dB for both sparse and dispersive AIRs. |
| | • The improvement of SC-IPNLMS compared to NLMS are 10 dB and 6 dB for sparse and dispersive AIRs, respectively. |

## V. CONCLUSION

We have presented various classification of echo cancellation algorithms.
Observations related to Proportionate-type algorithms and sparseness-controlled algorithms are as follows,

- Proportionate-type algorithms converge faster than the normalized least-mean-square (NLMS) algorithm, especially for sparse impulse responses.
- Among these, proportionate-type NLMS algorithms. the IPNLMS algorithm is one of the most attractive choices, mainly because of its robustness to echo paths with different sparseness degrees.
- NLMS algorithm is one of the most popular for AEC due to its straightforward implementation and low complexity compared to, other algorithms, for example, the Recursive Least Squares (RLS) algorithm.
- PNLMS (proportionate NLMS) [6] and MPNLMS (µ PNLMS ) [19] algorithms have been proposed for sparse system identification.
- MPNLMS ( µ PNLMS) algorithm is proposed to improve the convergence of PNLMS
- IPNLMS employs a combination of proportionate (PNLMS) and non-proportionate (NLMS) adaptation. IPNLMS [25] is a very good approximation of the filter, while being more convenient from a practical point of view.
- Simulation results for sparseness-controlled approach using white Gaussian noise (WGN) and speech input signals, show improved performance over existing methods. The proposed algorithms achieve these improvement with only a modest increase in computational complexity.
- The sparseness-controlled algorithms achieve fast convergence for both sparse and dispersive AIRs and are effective for AEC.
- In order to address the problem of slow convergence in PNLMS and MPNLMS for dispersive AIR, we require the step-size control elements to be robust to the sparseness of the impulse response.
- SC-IPNLMS improves the performance of the IPNLMS when the impulse response is sparse, the algorithm allocates more weight to the proportionate term

# REFERENCES

[1] Pradeep Loganathan, *et al*, A Class of Sparseness-Controlled Algorithms for Echo Cancellation, IEEE Transactions On Audio, Speech, And Language Processing, Vol. 17, No. 8, November 2009

[2] Jerónimo Arenas-García, *et al*, Adaptive Combination of Proportionate Filters for Sparse Echo Cancellation, IEEE Transactions On Audio, Speech, And Language Processing, Vol. 17, No. 6, August 2009

[3] Qin Li, *et al*, Design Of Oversampled Dft Modulated Filter Banks Optimized For Acoustic Echo Cancellation, Microsoft Research, ICASSP 2009

[4] Yekutiel Avargel, *et al*, Modeling and Identification of Nonlinear Systems in the Short-Time Fourier Transform Domain, IEEE Transactions On Signal Processing, Vol. 58, No. 1, January 2010

[5] Constantin Paleologu, *et al*, An Efficient Proportionate Affine Projection Algorithm for Echo Cancellation, IEEE Signal Processing Letters, Vol. 17, No. 2, February 2010

[6] D. L. Duttweiler, "Proportionate normalized least mean square adaptation in echo cancellers," IEEE Trans. Speech Audio Process., vol. 8, no. 5, pp. 508–518, Sep. 2000.

[7] A. W. H. Khong, J. Benesty, and P. A. Naylor, "Stereophonic acoustic echo cancellation: Analysis of the misalignment in the frequency domain," IEEE Signal Process. Lett., vol. 13, no. 1, pp. 33–36, Jan. 2006.

[8] H. Deng and M. Doroslovacki, "Wavelet-based MPNLMS adaptive algorithm for network echo cancellation," EURASIP J. Audio, Speech, Music Process., 2007.

[9] R. H. Kwong and E. Johnston, "A variable step size LMS algorithm," IEEE Trans. Signal Process., vol. 40, no. 7, pp. 1633–1642, Jul. 1992.

[10] J. Sanubari, "A new variable step size method for the LMS adaptive filter," in Proc. IEEE Asia-Pacific Conf. Circuits Syst., 2004, pp. 501–504.

[11] B. A. Schnaufer and W. K. Jenkins, "New data-reusing LMS algorithms for improved convergence," in Proc. 27th Asilomar Conf. Signals, Syst., Comput., 1993, pp. 1584–1588.

[12] K. A. G. Robert, A. Soni, and W. K. Jenkins, "Low-complexity data reusing methods in adaptive filtering," IEEE Trans. Signal Process., vol. 52, no. 2, pp. 394–405, Feb. 2004.

[13] A. W. H. Khong and P. A. Naylor, "Selective-tap adaptive algorithms in the solution of the non-uniqueness problem for stereophonic acoustic echo cancellation," IEEE Signal Processing Lett., vol. 12, no. 4, pp. 269–272, Apr. 2005.

[14] P. A. Naylor and A. W. H. Khong, "Affine projection and recursive least squares adaptive filters employing partial updates," in Proc. 38th Asilomar Conf. Signals, Syst. Comput., Nov. 2004, vol. 1, pp. 950–954.

[15] K. A. Lee and S. Gan, "Improving convergence of the NLMS algorithm using constrained subbands updates," IEEE Signal Process. Lett., vol. 11, no. 9, pp. 736–739, Sep. 2004.

[16] A. Deshpande and S. L. Grant, "A new multi-algorithm approach to sparse system adaptation," in Proc. Eur. Signal Process. Conf., 2005.

[17] S. L. Gay, "An efficient, fast converging adaptive filter for network echo cancellation," , vol. 1, pp. 394–398, Nov. 1998.

[18] J. Benesty and S. L. Gay, "An improved PNLMS algorithm," in Proc. IEEE Int. Conf. Acoust. Speech Signal Process., 2002, vol. 2, pp. 1881–1884.

[19] H. Deng and M. Doroslovacki, "Improving convergence of the PNLMS algorithm for sparse impulse response identification," IEEE Signal Process. Lett., vol. 12, no. 3, pp. 181–184, Mar. 2005.

[20] G. W. Elko, E. Diethorn, and T. Gänsler, "Room impulse response variation due to thermal fluctuation and its impact on acoustic echo cancellation," in Proc. Int. Workshop Acoust. Echo Noise Control, 2003, pp. 67–70.

[21] H. Sabine, "Room acoustics," Trans. IRE Professional Group Room Acoust., vol. 1, no. 4, pp. 4–12, Jul. 1953.

[22] S. Haykin, Adaptive Filter Theory. Upper Saddle River, NJ: Prentice-Hall, 2002.

[23] A. H. Sayed, Fundamentals of Adaptive Filtering. New York: Wiley, 2003.

[24] S. Makino, Y. Kaneda, and N. Koizumi, "Exponentially weighted stepsize NLMS adaptive filter based on the statistics of a room impulse response," IEEE Trans. Speech Audio Process., vol. 1, no. 1, pp. 101–108, Jan. 1993.

[25] J. Benesty and S. L. Gay, "An improved PNLMS algorithm," in Proc. ICASSP'02, Orlando, FL, 2002, vol. II, pp. 1881–1883.

[26] J. Arenas-García, M. Martínez-Ramón, A. Navia-Vázquez, and A. R. Figueiras-Vidal, "Plant identification via adaptive combination of transversal filters," Signal Process., vol. 86, pp. 2430–2438, 2006.

[27] J. Arenas-García, V. Gómez-Verdejo, and A. R. Figueiras-Vidal, "New algorithms for improved adaptive convex combination of LMS transversal filers," IEEE. Trans. Instrum. Meas., vol. 54, no. 6, pp. 2239–2249, Dec. 2005.

[28] J. Arenas-García, A. R. Figueiras-Vidal, and A. H. Sayed, "Meansquare performance of a convex combination of two adaptive filters," IEEE Trans. Signal Process., vol. 54, no. 3, pp. 1078–1090, Mar. 2006.

[29] E. Hansler and G. Schmidt, Acoustic Echo and Noise Control: A Practical Approach, Chapter 9, New York: Wiley-IEEE, 2004.

[30] A. Gilloire and M. Vetterli, "Adaptive filtering in subbands with critical sampling: analysis, experiments, and application to acoustic echo cancellation," IEEE Trans. on Signal Processing, vol. 40, no. 8, pp. 1862–1875, Aug. 1992.

[31] A. Guérin, G. Faucon, and R. L. Bouquin-Jeannés, "Nonlinear acoustic echo cancellation based on Volterra filters," IEEE Trans. Speech Audio Process., vol. 11, no. 6, pp. 672–683, Nov. 2003.

[32] G. O. Glentis, P. Koukoulas, and N. Kalouptsidis, "Efficient algorithms for Volterra system identification," IEEE Trans. Signal Process., vol. 47, no. 11, pp. 3042–3057, Nov. 1999.

[33] R. D. Nowak, "Penalized least squares estimation of Volterra filters and higher order statistics," IEEE Trans. Signal Process., vol. 46, no. 2, pp. 419–428, Feb. 1998.

[34] S. Im and E. J. Powers, "A block LMS algorithm for third-order frequency- domain Volterra filters," IEEE Signal Process. Lett., vol. 4, no. 3, pp. 75–78, Mar. 1997.

[35] K. I. Kim and E. J. Powers, "A digital method of modelling quadratically nonlinear systems with a general random input," IEEE Trans. Acoust., Speech, Signal Process., vol. 36, no. 11, pp. 1758–1769, Nov.1988.

[36] C. H. Tseng and E. J. Powers, "Batch and adaptive Volterra filtering of cubically nonlinear systems with a Gaussian input," in Proc. IEEE Int Symp. Circuits Syst. (ISCAS), 1993, vol. 1, pp. 40–43.

[37] J. Benesty and Y. Huang, Eds., Adaptive Signal Processing–Applications to Real-World Problems. Berlin, Germany, Springer-Verlag, 2003.

[38] J. Benesty and S. L. Gay, "An improved PNLMS algorithm," in Proc. IEEE ICASSP, 2002, pp. II-1881–II-1884.